\documentclass[12pt]{article}
\usepackage{amssymb}
\usepackage{amsfonts}
\usepackage{amsmath}

\newtheorem{theorem}{Theorem}

\newtheorem{corollary}{Corollary}

\newtheorem{lemma}{Lemma}

\newtheorem{proposition}{Proposition}
\newtheorem{remark}{Remark}

\begin{document}

\title{On extensions of partial priorities in school choice\thanks{%
We would like to thank Onur Kesten for helpful comments. This work was
supported by JSPS KAKENHI Grant Numbers 20K01675 and 22K01402.}}
\author{Minoru Kitahara \and Yasunori Okumura\thanks{%
Corresponding author Address: 2-1-6, Etchujima, Koto-ku, Tokyo, 135-8533
Japan. Phone:+81-3-5245-7300. Fax:+81-3-5245-7300. E-mail:
okuyasu@gs.econ.keio.ac.jp}}
\maketitle

\begin{center}
\textbf{Abstract}
\end{center}

We consider a school choice matching model where priorities for schools are
represented by binary relations that may not be total orders. Even in that
case, it is necessary to construct total orders from the priority relations
to execute several mechanisms. We focus on the extensions of the priority
relations, because a matching that is stable for any extension profile is
also stable for the profile of priority relations. We introduce a class of
algorithms for deriving one of the extensions of acyclic priority relations
and characterize them utilizing the class. We show that if the priorities
are partial orders, then for each stable matching for the profile of
priority relations, an extension profile for which it is also stable exists.
Furthermore, if multiple stable matchings ranked by Pareto dominance exist,
there is an extension for which all of these matchings are stable. We
provide several applications of these results.



\newpage

\section{Introduction}

In many school choice systems, the priorities for schools of students are
determined by laws and/or government policies, rather than reflecting the
schools' preferences. Therefore, these priorities should be represented by
binary relations (hereafter priority relations) which may not be total
orders; that is, may not be complete and/or transitive. This is because,
preserving unnecessary priorities would cause efficiency loss. See, for
example, Erdil and Ergin (2008), Abdulkadiro\u{g}lu et al. (2009), Kesten
(2010), Dur et al. (2019) and Kitahara and Okumura (2021) on the facts in
detail.

However, even if a priority relation for a school is not a total order, we
often need to construct a total order from the priority relation to execute
several influential mechanisms such as the deferred acceptance (hereafter
DA) mechanisms (Gale and Shapley (1962)) and the efficiency adjusted DA
mechanisms (hereafter EADAMs) (Kesten (2010)). The results of the mechanisms
are dependent on the total orders constructed from the priority relations
for schools. In particular, Erdil and Ergin (2008) show that a result of the
DA mechanism with a profile of tie-broken total orders is stable for the
profile of priority relations but may be Pareto dominated by another stable
matching for it.

We focus on a (total order) extension of a priority relation, which is a
total order that contains the priority relation. In particular, we introduce
a class of algorithms to derive one of the extensions of an acyclic priority
relation, and characterize them by utilizing this class. If the priority
relations are acyclic, then we can obtain a stable matching for the profile
of priority relations by using the extension constructed by any of the
algorithms in the class and a DA mechanism. Furthermore, this result implies
the existence of stable matchings.

Next, we focus on the case where the priority relation for each school is a
partial order (which is transitive) and stable matchings for the profile of
partial orders. For more information on school choice models with partial
order priorities, see Dur et al. (2019) and Kitahara and Okumura (2021). We
show that for any stable matching for any partial orders profile, there
exists its extension profile for which the matching is stable. Furthermore,
if there are multiple stable matchings for the profile that are ranked by
Pareto dominance, then there is its extension profile for which \textit{all}
of those matchings are stable.

We provide several applications of our results. First, we generalize results
of Erdil and Ergin (2008) on the set of stable matchings and the set of
student optimal stable matchings, which are stable matchings that are not
Pareto dominated by any stable matching. Specifically, we show that the set
of stable matchings for a partial order profile is equivalent to the union
of the sets of stable matchings for every extension profile.$\mathcal{\ }$%
Moreover, we show that any student optimal stable matching can be obtained
by using the student proposing DA mechanism with an extension profile.

Second, we revisit tiebreaking rules. Abdulkadiro\u{g}lu et al. (2009)
provide a theoretical result that supports single tiebreaking rules, in
which the manners of tiebreaking are the same across all schools, as proper
tiebreaking rules that are used with the student proposing DA mechanism,
when the priority relations are weak orders. However, we show that if the
priority relation of a school is not weak order, then the theoretical result
of Abdulkadiro\u{g}lu et al. (2009) is not continued to hold. Therefore, in
that case, the single tiebreaking rules are not supported theoretically.

Third, we focus on EADAMs, which is recently discussed by many previous
studies.\footnote{%
Moreover, in 2019, the Flemish Ministry of Education undertook the first
attempt to implement EADAM in the school choice system in Flanders (Cerrone
et al. (2022)).} See Cerrone et al. (2022) for the survey. We consider a
variant of EADAMs that is used to derive a student optimal stable matching
for the profile of priority relations, when the priority relations for some
schools are not total orders. We show that if the priority relations are
partial orders, then any student optimal stable matching for the profile of
the priority relations is derived by a mechanism in the EADAMs class.

Fourth, we examine the school choice model with allowable priority
violations introduced by Kesten (2010) and generalized by Dur et al. (2019).
We show that the general model with allowable priority violations is
essentially equivalent to the usual school choice model with acyclic
priority relations, which may not be partial orders. By the existence result
above, this implies the existence of matchings in the general model.

\section{Model and Results}

Let $B$ be a binary relation on a set $X$ that is asymmetric; that is, for
any $x,y\in X,$ $\left( x,y\right) \in B$ implies $\left( y,x\right) \notin B
$. Here, we basically follow Fishburn (1970). A binary relation on a set $X$
denoted by $B$ is

\begin{description}
\item \textbf{complete}\textit{\ }if\textit{\ }$x\neq y$ implies\textit{\ }$%
\left( x,y\right) \in B$ or $\left( y,x\right) \in B$ throughout $X$,

\item \textbf{negatively} \textbf{transitive}\textit{\ }if [$\left(
x,y\right) \notin B$ and $\left( y,z\right) \notin B$] implies $\left(
x,z\right) \notin B$, for all $x,y,z\in X$,

\item \textbf{transitive} if [$\left( x,y\right) \in B$ and $\left(
y,z\right) \in B$] implies $\left( x,z\right) \in B$, for all $x,y,z\in X,$

\item \textbf{acyclic }if for all $K\in \left\{ 2,3,\cdots \right\} $ and
for all $x_{0},x_{1},\cdots ,x_{K}\in X$, $\left( x_{k-1},x_{k}\right) \in B$
and $\left( x_{k},x_{k-1}\right) \notin B$ for all $k\in \left\{ 1,\cdots
,K\right\} $ implies $\left( x_{K},x_{0}\right) \notin B$.
\end{description}

Since we focus only on asymmetric binary relation, if $B$ is transitive,
then it is acyclic.

An asymmetric binary relation on $X$ is called a (strict) \textbf{partial
order} on $X$ if it is transitive. A partial order on $X$ is called a 
\textbf{weak order} on $X$ if it is negatively transitive. Moreover, a weak
order on $X$ is called a \textbf{total order} if it is complete. If $B$ is a
total order, we write $B:$ $x_{1}$ $x_{2}$ $x_{3}$ $\cdots ,$ meaning that $%
\left( x_{t},x_{t^{\prime }}\right) \in B$ for all $t<t^{\prime }$.\ 

Let $I$ and $S$ be the finite sets of students with $\left\vert I\right\vert
\geq 3$ and schools, respectively. Each student $i\in I$ has a total order
on $S\cup \left\{ \emptyset \right\} $ denoted by $P_{i}$, where $%
sP_{i}s^{\prime }$ means that $i$ prefers $s\in S\cup \left\{ \emptyset
\right\} $ to $s^{\prime }\in S\cup \left\{ \emptyset \right\} $ and $%
\emptyset $ represents her/his best outside option. Let $sR_{i}s^{\prime }$
mean $sP_{i}s^{\prime }$ or $s=s^{\prime }$.

Each school $s$ has a capacity constraint represented by $q_{s}\in \mathbb{Z}%
_{++}$ and $q=\left( q_{s}\right) _{s\in S}.$ Let $\succ _{s}$ be a \textbf{%
priority relation} \textbf{for school} $s$ that is an asymmetric binary
relation on $I$, where $\left( i,j\right) \in \succ _{s}$ means that $i$ has
a higher priority than $j$ for school $s$.\footnote{%
In the previous study, a priority relation is called a \textquotedblleft
priority order\textquotedblright , because, contrary to this study, it is at
least a partial order.} Let $\mathcal{A},$ $\mathcal{P},$ $\mathcal{W}$ and $%
\mathcal{T}$ be the set of all possible acyclic, partial order, weak order
and total order priorities, respectively. Note that $\mathcal{T\subsetneq
W\subsetneq P\subsetneq A}$. Let $\succ =\left( \succ _{s}\right) _{s\in S}$
be a priority profile.

Let $G=\left( I,S,P,\succ ,q\right) $ be a \textbf{school choice problem}.
In this study, since we fix $\left( I,S,P,q\right) $, a school choice
problem is simply denoted by $G=\succ $.

A \textbf{matching} $\mu $ is a mapping satisfying $\mu (i)\in S\cup
\{\emptyset \},$ $\mu \left( s\right) \subseteq I$, and $\mu (i)=s$ if and
only if $i\in \mu \left( s\right) $. $\mu (i)=\emptyset $ means that $i$ is
unmatched and $\mu (i)=s\in S$ means that $i$ is matched to $s$ under $\mu $.

A matching $\mu $ is \textbf{individually rational} if $\mu \left( i\right)
R_{i}\emptyset $ for all $i\in I$. A matching $\mu $ is \textbf{non-wasteful}
if $sP_{i}\mu \left( i\right) $ implies $\left\vert \mu \left( s\right)
\right\vert =q_{s}$ for all $i\in I$ and all $s\in S$. A matching $\mu $ 
\textbf{violates} the priority of $i\notin \mu \left( s\right) $ for $s$
over $j\in \mu \left( s\right) $ if $sR_{i}\mu \left( i\right) \ $and $%
\left( i,j\right) \in \succ _{s}$. If a matching $\mu $ does not violate any
priorities for\textbf{\ }$\succ _{s}$ for all $s\in S$, then it is \textbf{%
fair }for $\succ $. A matching $\mu $ is \textbf{stable} for $\succ $ if it
is individually rational, non-wasteful and fair for $\succ $.

A matching $\mu $ \textbf{is} \textbf{Pareto dominated} \textbf{by} $\mu
^{\prime }$ if $\mu ^{\prime }\left( i\right) R_{i}\mu \left( i\right) $ for
all $i\in I$ and $\mu ^{\prime }\left( i\right) P_{i}\mu \left( i\right) $\
for some $i\in I$. A matching is a \textbf{student\ optimal stable matching }%
(hereafter\textbf{\ SOSM}) for $\succ $ if it is stable for $\succ $ and is
not Pareto dominated by any stable matching for $\succ $.

Following Erdil and Ergin (2008), let $\mathcal{S}^{\succ }$ be a set of all
stable matchings for $\succ $ and $f^{\succ }\left( \subseteq \mathcal{S}%
^{\succ }\right) $ be a set of all SOSMs for $\succ $. Gale and Shapley
(1962) show that if $\succ \in \mathcal{T}^{\left\vert S\right\vert }$, then 
$f^{\succ }$ is a singleton; that is, $\mu ^{\ast }\in f^{\succ }$ Pareto
dominates any $\mu \in \mathcal{S}^{\succ }\setminus \left\{ \mu ^{\ast
}\right\} $. Moreover, in that case, $f^{\succ }$ is derived via the
(student proposing) deferred acceptance algorithm (hereafter the DA
algorithm).

A binary relation $\hat{\succ}_{s}$ is an \textbf{extension} of $\succ _{s}$
if $\succ _{s}\subseteq \hat{\succ}_{s}$. We immediately have the following
result.

\begin{lemma}
Suppose that $\hat{\succ}_{s}$ is an extension of $\succ _{s}$ for all $s\in
S$. If a matching is stable for $\hat{\succ}=\left( \hat{\succ}_{s}\right)
_{s\in S}$, then it is also stable for $\succ $.
\end{lemma}

Further, a binary relation $\hat{\succ}_{s}$ is a \textbf{total order} 
\textbf{extension} of $\succ _{s}$ if $\succ _{s}\subseteq \hat{\succ}_{s}$
and $\hat{\succ}_{s}\in \mathcal{T}$. Hereafter, since we focus only on a
total order extension, we simply write an \textquotedblleft
extension\textquotedblright\ representing a \textquotedblleft total order
extension\textquotedblright .

We introduce a method to construct an extension of $\succ _{s}$. First, for $%
I^{\prime }\subseteq I$ and $\succ _{s}$, let $M\left( I^{\prime },\succ
_{s}\right) $ be the \textbf{maximal set}; that is, 
\begin{equation*}
M\left( I^{\prime },\succ _{s}\right) =\left\{ i\in I^{\prime }\text{ }%
\left\vert \text{ }\left( j,i\right) \notin \succ _{s}\text{ for all }j\in
I^{\prime }\setminus \left\{ i\right\} \right. \right\} ,
\end{equation*}%
which is not empty if $\succ _{s}\in \mathcal{A}$ (Bossert and Suzumura
(2010)).

We consider the following class of algorithms, which is discussed by Okumura
(2023).\newline

\textbf{Step} $1$: Let $i_{1}\in M\left( I,\succ _{s}\right) $.

\textbf{Step} $t=2,\cdots ,\left\vert I\right\vert $: Let $i_{t}\in M\left(
I\setminus \left\{ i_{1},\cdots ,i_{t-1}\right\} ,\succ _{s}\right) $.

Let $\succ _{s}^{\prime }:$ $i_{1}$ $i_{2}$ $\cdots $ $i_{\left\vert
I\right\vert }$.\newline

Note that since $M\left( I^{\prime },\succ _{s}\right) $ may not be
singleton, this is a class of algorithms. For each $\succ _{s}\in \mathcal{A}%
,$ we call this class as the \textbf{sequential maximal ordering} (hereafter 
\textbf{SMO}) class.

\begin{proposition}
A priority $\succ _{s}$ has its extension if and only if $\succ _{s}\in 
\mathcal{A}$. For any $\succ _{s}\in \mathcal{A},$ $\succ _{s}^{\prime }$ is
an extension of $\succ _{s}$ if and only if it is obtained by an algorithm
within the SMO class for $\succ _{s}$.
\end{proposition}

\textbf{Proof.} We use a result of Okumura (2023). In his terminology, for a
binary relation $B$ on $X$, a total order $B^{\prime }$ is said to be 
\textbf{consistent to }$B$ if for all $a,b\in X$, $\left( a,b\right) \in
P\left( B\right) $ implies $\left( a,b\right) \in B^{\prime }$.

\begin{remark}
Okumura (2023, Proposition 1) The set of total orders that are consistent to 
$B$ is not empty if and only if $B$ is acyclic. Moreover, a total order is
consistent to $B$ if and only if it is obtained by an algorithm within the
SMO class for $B$.
\end{remark}

If $B$ is asymmetric, then $\left( a,b\right) \in P\left( B\right) $ if and
only if $\left( a,b\right) \in B$. Since any priority $\succ _{s}$ is
asymmetric in this model, a total order is consistent to $\succ _{s}$ if and
only if it is an extension of $\succ _{s}$. By this fact and Remark 1, we
have Proposition 1. \textbf{Q.E.D.}\newline

Let $E\left( \succ _{s}\right) $ be the set of extensions of $\succ _{s}$
and $\mathcal{E}\left( \succ \right) $ be the set of the extension profiles;
that is, $\succ ^{\prime }=\left( \succ _{s}^{\prime }\right) _{s\in S}\in 
\mathcal{E}\left( \succ \right) $ implies $\succ _{s}^{\prime }\in E\left(
\succ _{s}\right) $ for all $s\in S$. Proposition 1 implies that for any $%
\succ _{s}\in \mathcal{A},$ if $i_{t}$ is randomly chosen from $M\left(
I\setminus \left\{ i_{1},\cdots ,i_{t-1}\right\} ,\succ _{s}\right) $ for
each step $t$ of an SMO, then each profile in $\mathcal{E}\left( \succ
\right) $ is realized with a positive probability.

We have the following existence result, which is a generalization of a
result of Kitahara and Okumura (2021, Proposition 1).

\begin{corollary}
If $\succ \in \mathcal{A}^{\left\vert S\right\vert }$, then $f^{\succ
}\left( \subseteq \mathcal{S}^{\succ }\right) $ is nonempty.
\end{corollary}

\textbf{Proof. }By Proposition 1, for any $\succ \in \mathcal{A}^{\left\vert
S\right\vert }$, there is $\succ ^{\prime }\in \mathcal{E}\left( \succ
\right) $. By the result of Gale and Shapley, there is a stable matching for 
$\succ ^{\prime }$. Therefore, by Lemma 1, the stable matching for $\succ
^{\prime }$ is also stable for $\succ $. \textbf{Q.E.D.}\newline

We introduce our main result.

\begin{theorem}
Suppose that $\succ \in \mathcal{P}^{\left\vert S\right\vert }$ and there
are $K(\geq 1)$ stable matchings for $\succ $ denoted by $\mu _{1},\cdots
,\mu _{K}$ such that $\mu _{k^{\prime }}$ Pareto dominates $\mu _{k}$ for
all $1\leq k<k^{\prime }\leq K$. Then, there is $\succ ^{\ast }\in \mathcal{E%
}\left( \succ \right) $ such that all $\mu _{1},\cdots ,\mu _{K}$ are stable
for $\succ ^{\ast }$.
\end{theorem}

\textbf{Proof. }For each $k\in \left\{ 1,\cdots ,K\right\} $ and for each $%
s\in S$, let 
\begin{equation*}
A_{s}^{k}=\left\{ \left. \left( i,j\right) \text{ }\right\vert \text{ }\mu
_{k}\left( i\right) =sP_{j}\mu _{k}\left( j\right) \right\} .
\end{equation*}%
and

\begin{equation*}
\succ _{s}^{\prime }=\succ _{s}\bigcup \left( \bigcup\limits_{k\in \left\{
1,\cdots ,K\right\} }A_{s}^{k}\right) \text{.}
\end{equation*}%
As mentioned by Erdil and Ergin (2008, p675), $\mu _{k}$ is stable for the
profile of extensions of $\succ _{s}\cup A_{s}^{k}$ for all $s\in S$.%
\footnote{%
One of the authors (Okumura (2014)) had missed this fact and realized the
mistake thanks to the comment of the other author (Kitahara).} Thus, if
there exists an extension of $\succ _{s}^{\prime }$ for all $s\in S,$ then
all $\mu _{1},\cdots ,\mu _{K}$ are stable for that profile of extension. By
Proposition 1, the following result is sufficient for the proof of its
existence.

\begin{lemma}
Suppose that $\succ \in \mathcal{P}^{\left\vert S\right\vert }$ and there
are $K(\geq 1)$ stable matchings for $\succ $ denoted by $\mu _{1},\cdots
,\mu _{K}$ such that $\mu _{k^{\prime }}$ Pareto dominates $\mu _{k}$ for
all $1\leq k<k^{\prime }\leq K$. Then, $\succ ^{\prime }\in \mathcal{A}%
^{\left\vert S\right\vert }$.
\end{lemma}

\textbf{Proof.} First, we show that $\succ _{s}^{\prime }$ is asymmetric;
that is, we let $\left( i,j\right) \in \succ _{s}^{\prime }$ and show that $%
\left( j,i\right) \notin \succ _{s}^{\prime }$. First, suppose $\left(
i,j\right) \in \succ _{s}$. Since $\mu _{k}$ is fair for $\succ $, $\mu
_{k}\left( j\right) =sP_{i}\mu _{k}\left( i\right) $ is not satisfied and
thus $\left( j,i\right) \notin A_{s}^{k}$ for any $k=1,\cdots ,K$. Second,
suppose $\left( i,j\right) \in A_{s}^{k}$ for some $k=1,\cdots ,K$. By the
fact above, $\left( j,i\right) \notin \succ _{s}$. Thus, toward a
contradiction, suppose $\left( j,i\right) \in A_{s}^{k^{\prime }}$. Since $%
\left( i,j\right) \in A_{s}^{k}$ and $\mu _{k}$ is stable, $\left(
j,i\right) \notin \succ _{s}$. Then, $\mu _{k^{\prime }}$ either Pareto
dominates or is Pareto dominated by $\mu _{k}$. In the former case, $\left(
j,i\right) \in A_{s}^{k^{\prime }}$ and $\left( i,j\right) \in A_{s}^{k}$
imply $\mu _{k^{\prime }}\left( j\right) =sP_{i}\mu _{k^{\prime }}\left(
i\right) R_{i}\mu _{k}\left( i\right) =s,$ which is a contradiction. In the
latter case, $\left( j,i\right) \in A_{s}^{k^{\prime }}$ and $\left(
i,j\right) \in A_{s}^{k}$ imply $\mu _{k}\left( i\right) =sP_{j}\mu
_{k}\left( j\right) R_{j}\mu _{k^{\prime }}\left( j\right) =s,$ which is
also a contradiction. Thus, $\succ _{s}^{\prime }$ is asymmetric.

Second, we show $\succ _{s}^{\prime }$ is acyclic. Suppose not; that is, $%
\succ _{s}^{\prime }$ has a cycle $\left( i_{0},i_{1},i_{2},\cdots
,i_{D}\right) $ of distinct students such that $\left( i_{d-1},i_{d}\right)
\in \succ _{s}^{\prime }$ for all $d\in \left\{ 1,\cdots ,D\right\} $ and $%
\left( i_{D},i_{0}\right) \in \succ _{s}^{\prime }$. Without loss of
generality, we assume this cycle is a shortest one. Since $\succ _{s}$ is
acyclic, there is $\left( k,d\right) \in \left\{ 1,\cdots ,K\right\} \times
\left\{ 1,\cdots ,D\right\} $ such that $\left( i_{d-1},i_{d}\right) \in
A_{s}^{k}$. Without loss of generality, suppose $\left( i_{0},i_{1}\right)
\in A_{s}^{k}$. Then, there is no other $d=2,\cdots ,D$ such that $\left(
i_{d-1},i_{d}\right) \in A_{s}^{k}$. We show this fact. Suppose not; that
is, $\left( i_{d-1},i_{d}\right) \in A_{s}^{k}$ for some $d=2,\cdots ,D$.
Then $\left( i_{0},i_{d}\right) \in A_{s}^{k}\subseteq \succ _{s}^{\prime }$
and therefore there must be a shorter cycle $\left( i_{0},i_{d},\cdots
,i_{D}\right) $ contradicting the shortest cycle assumption.

Next, we show that there is $\left( k^{\prime },d\right) \in \left\{
1,\cdots ,K\right\} \setminus \left\{ k\right\} \times \left\{ 2,\cdots
,D\right\} $ such that $\left( i_{d-1},i_{d}\right) \in A_{s}^{k^{\prime }}$%
. Suppose not; that is, there is no such $\left( k^{\prime },d\right) $.
Then, $\left( i_{d-1},i_{d}\right) \in \succ _{s}$ for all $d=2,\cdots ,D$
and thus $\left( i_{1},i_{2}\right) \in \succ _{s}$ and $\left(
i_{2},i_{3}\right) \in \succ _{s}$. By the transitivity of $\succ _{s},$ $%
\left( i_{1},i_{3}\right) \in \succ _{s}$ and thus there must be a shorter
cycle $\left( i_{0},i_{1},i_{3},\cdots ,i_{D}\right) $ contradicting the
shortest cycle assumption. Therefore, there is $\left( k^{\prime },d\right)
\in \left\{ 1,\cdots ,K\right\} \setminus \left\{ k\right\} \times \left\{
2,\cdots ,D\right\} $ such that $\left( i_{d-1},i_{d}\right) \in
A_{s}^{k^{\prime }}$.

First, suppose $k^{\prime }>k$. Then, since $\mu _{k^{\prime }}$ Pareto
dominates $\mu _{k}$, $\mu _{k^{\prime }}\left( i_{d}\right) R_{i_{d}}\mu
_{k}\left( i_{d}\right) $. Moreover, since $\mu _{k}\left( i_{0}\right) =s$, 
$\left( i_{0},i_{d}\right) \in A_{s}^{k}\subseteq \succ _{s}^{\prime }$
contradicting the shortest cycle assumption. Second, suppose $k>k^{\prime }$%
. Then, $\left( i_{0},i_{1}\right) \in A_{s}^{k}$ and $\left(
i_{d-1},i_{d}\right) \in A_{s}^{k^{\prime }}$. By the definitions of $%
A_{s}^{k^{\prime }}$ and $A_{s}^{k},$ $\mu _{k^{\prime }}\left(
i_{d-1}\right) =s$ and $sP_{i_{1}}\mu _{k}\left( i_{1}\right) $. Since $\mu
_{k}$ Pareto dominates $\mu _{k^{\prime }}$, $\mu _{k}\left( i_{1}\right)
R_{i_{1}}\mu _{k^{\prime }}\left( i_{1}\right) $. Therefore, $s=\mu
_{k^{\prime }}\left( i_{d-1}\right) P_{i_{1}}\mu _{k^{\prime }}\left(
i_{1}\right) $, which implies $\left( i_{d-1},i_{1}\right) \in
A_{s}^{k}\subseteq \succ _{s}^{\prime }$. However, there must be a shorter
cycle $\left( i_{1},i_{2},\cdots ,i_{d-1}\right) $ contradicting the
shortest cycle assumption. \textbf{Q.E.D.}\newline

We provide a counterexample of Theorem 1 (and Lemma 2) when $\succ _{s}\in 
\mathcal{A\setminus P}$ for some $s\in S$.\newline

\textbf{Example 1.} Let 
\begin{gather*}
\succ _{s}=\left\{ \left( i_{1},i_{2}\right) ,\left( i_{2},i_{0}\right)
\right\} , \\
\succ _{s^{\prime }}:i_{2}\text{ }i_{1}\text{ }i_{0},\text{ }%
q_{s}=q_{s^{\prime }}=1, \\
P_{i_{0}}:s\text{ }s^{\prime },\text{ }P_{i_{1}}:s\text{ }s^{\prime },\text{ 
}P_{i_{2}}:s^{\prime }\text{ }s.
\end{gather*}%
Since $\left( i_{1},i_{0}\right) \notin \succ _{s}$, $\succ _{s}$ is not
transitive. Let $\mu $ be such that $\mu \left( i_{0}\right) =s$ and $\mu
\left( i_{2}\right) =s^{\prime }$. Then, $\mu $ is an SOSM for $\succ $.
However, 
\begin{equation*}
\succ _{s}\cup \left\{ \left. \left( i,j\right) \text{ }\right\vert \text{ }%
\mu \left( i\right) =sP_{j}\mu \left( j\right) \right\} =\left\{ \left(
i_{0},i_{1}\right) ,\left( i_{1},i_{2}\right) ,\left( i_{2},i_{0}\right)
\right\}
\end{equation*}%
is cyclic. Therefore, $\mu $ is not stable for any $\succ ^{\prime }\in 
\mathcal{E}\left( \succ \right) $.\newline

Hence the transitivity of priorities is important for Theorem 1 and Lemma 2.

\section{Applications}

\subsection{Sets of stable matchings and SOSMs}

By Theorem 1, we have the following results, which are generalizations of
Observations 1 and 2 of Erdil and Ergin (2008) that focus only on the case
where $\succ \in \mathcal{W}^{\left\vert S\right\vert }$.

\begin{corollary}
For any $\succ \in \mathcal{P}^{\left\vert S\right\vert }$, 
\begin{equation*}
\mathcal{S}^{\succ }=\bigcup\limits_{\succ ^{\prime }\in \mathcal{E}\left(
\succ \right) }\mathcal{S}^{\succ ^{\prime }}\text{ and }f^{\succ }\subseteq
\bigcup\limits_{\succ ^{\prime }\in \mathcal{E}\left( \succ \right)
}f^{\succ ^{\prime }}.
\end{equation*}
\end{corollary}

Note that Example 1 implies that Corollary 2 does not hold when $\succ
_{s}\in \mathcal{A\setminus P}$ for some $s\in S$. The latter result implies
that any SOSM for $\succ \in \mathcal{P}^{\left\vert S\right\vert }$ can be
obtained by using the DA algorithm with some $\succ ^{\prime }\in \mathcal{E}%
\left( \succ \right) $.

\subsection{Tiebreaking}

We consider a deterministic tiebreaking rule \`{a} la Abdulkadiro\u{g}lu et
al. (2009). Let $r_{s}:I\rightarrow \left\{ 1,\cdots ,\left\vert
I\right\vert \right\} $ be a bijection for $s\in S$ and $\tau =\left(
r_{s}\right) _{s\in S}$. We define a $\tau $ \textbf{tiebreaking} \textbf{%
SMO algorithm }such that in Step $t=1,\cdots ,\left\vert I\right\vert $ of
SMO, $i_{t}$ is chosen from $M\left( I\setminus \left\{ i_{1},\cdots
,i_{t-1}\right\} ,\succ _{s}\right) $ if $r_{s}\left( i_{t}\right) \leq
r_{s}\left( i\right) $ for all $i\in M\left( I\setminus \left\{ i_{1},\cdots
,i_{t-1}\right\} ,\succ _{s}\right) $.

We consider a class of mechanisms such that, we first derive $\succ ^{\prime
}\in \mathcal{E}\left( \succ \right) $ via a tiebreaking SMO algorithm, and
second derive an SOSM for $\succ ^{\prime }$ via the DA algorithm.\ Erdil
and Ergin (2008) show that a mechanism in this class may fail to have any
SOSM for $\succ $, because for some $\succ \in \mathcal{W}^{\left\vert
S\right\vert }$ and some $\succ ^{\prime }\in \mathcal{E}\left( \succ
\right) $, any SOSM for $\succ ^{\prime }$ is Pareto dominated by a stable
matching for $\succ $. We show that even in the case of $\succ \in \mathcal{P%
}^{\left\vert S\right\vert }$, the failure of this mechanism is caused by an
improper tiebreaking rule.

\begin{corollary}
For all $\succ \in \mathcal{P}^{\left\vert S\right\vert }$, there is some $%
\succ ^{\prime }\in \mathcal{E}\left( \succ \right) $ such that $f^{\succ
^{\prime }}\subseteq f^{\succ }$.
\end{corollary}

This implies that if only we properly choose a tiebreaking rule, then we can
obtain an SOSM for $\succ $ via a mechanism in this class.

Next, we consider a single tiebreaking rule. If $r_{s}=r_{s^{\prime }}$ for
all $s,s^{\prime }\in S$, then $\tau $ is called a \textbf{single (common)
tiebreaking rule}. On the other hand, $\tau $ is called a \textbf{multiple
tiebreaking rule }if\textbf{\ }$r_{s}$ and $r_{s^{\prime }}$ may differ for $%
s,s^{\prime }\in S$. Let $\mathcal{E}^{c}\left( \succ \right) $ and $%
\mathcal{E}^{m}\left( \succ \right) $ be the sets of the extensions of $%
\succ $ obtained by the single tiebreaking rules and the multiple
tiebreaking rules, respectively. Note that $\mathcal{E}^{c}\left( \succ
\right) \subseteq \mathcal{E}^{m}\left( \succ \right) $ and, by Proposition
1, $\mathcal{E}^{m}\left( \succ \right) =\mathcal{E}\left( \succ \right) $
for any $\succ \in \mathcal{A}^{\left\vert S\right\vert }$.

The following result is introduced by Abdulkadiro\u{g}lu et al. (2009,
Proposition 2).

\begin{remark}
Suppose that $\succ \in \mathcal{W}^{\left\vert S\right\vert }$ and there is 
$\mu $ such that $\mu \in f^{\succ ^{\prime }}$ for some $\succ ^{\prime
}\in \mathcal{E}^{m}\left( \succ \right) ,$ but $\mu \in \mathcal{S}^{\succ
^{\prime \prime }}\setminus f^{\succ ^{\prime \prime }}$ for any $\succ
^{\prime \prime }\in \mathcal{E}^{c}\left( \succ \right) $. Then, $\mu
\notin f^{\succ }$.
\end{remark}

By Remark 2 and Theorem 1, we have the following result.

\begin{corollary}
If $\succ \in \mathcal{W}^{\left\vert S\right\vert }$, then%
\begin{equation*}
f^{\succ }\subseteq \bigcup\limits_{\succ ^{\prime }\in \mathcal{E}%
^{c}\left( \succ \right) }f^{\succ ^{\prime }}.
\end{equation*}
\end{corollary}

This result implies that if the priorities for all schools are weak orders,
then any SOSM can be obtained via the DA algorithm and the $\tau $
tiebreaking SMO algorithm where $\tau $ is a single tiebreaking rule.

On the other hand, in the next example, we show that this result is not
satisfied if $\succ _{s}\in \mathcal{P\setminus W}$ for some $s\in S$.

\textbf{Example 2. }Let $I=\left\{ i_{1},i_{2},i_{3},i_{4}\right\} $ and $%
S=\left\{ s_{1},s_{2},s_{3}\right\} $. Suppose\textbf{\ }%
\begin{gather*}
P_{i_{1}}:s_{3}\text{ }s_{1}\text{ }\emptyset ,\text{ }P_{i_{2}}:s_{1}\text{ 
}s_{2}\text{ }\emptyset ,\text{ }P_{i_{3}}:s_{2}\text{ }s_{3}\text{ }%
\emptyset ,\text{ }P_{i_{4}}:s_{3}\text{ }\emptyset , \\
\succ _{s_{1}}=\emptyset ,\succ _{s_{2}}=\emptyset ,\succ _{s_{3}}=\left\{
\left( i_{4},i_{1}\right) \right\} ,
\end{gather*}%
Since $\left( i_{4},i_{3}\right) \notin \succ _{s_{3}}$ and $\left(
i_{3},i_{1}\right) \notin \succ _{s_{3}},$ but $\left( i_{4},i_{1}\right)
\in \succ _{s_{3}},$ $\succ _{s_{3}}$ is not negatively transitive. The
matching $\mu $ such that $\mu \left( i_{1}\right) =s_{1}$, $\mu \left(
i_{2}\right) =s_{2},$ $\mu \left( i_{3}\right) =s_{3},$ and $\mu \left(
i_{4}\right) =\emptyset $ is an SOSM for $\succ $. Then, $f^{\succ ^{\prime
}}=\left\{ \mu \right\} $ where $\succ ^{\prime }\in \mathcal{E}\left( \succ
\right) $ if and only if 
\begin{equation*}
\left( i_{1},i_{2}\right) \in \succ _{s_{1}}^{\prime },\left(
i_{2},i_{3}\right) \in \succ _{s_{2}}^{\prime }\text{ and }\left(
i_{3},i_{1}\right) ,\left( i_{3},i_{4}\right) \in \succ _{s_{3}}^{\prime }.
\end{equation*}%
This implies $\succ ^{\prime }\notin \mathcal{E}^{c}\left( \succ \right) $,
because $r_{s_{1}}(i_{1})<r_{s_{1}}(i_{2})$, $%
r_{s_{2}}(i_{2})<r_{s_{2}}(i_{3})$ and $r_{s_{3}}(i_{3})<r_{s_{3}}(i_{1})$
must be satisfied. Therefore, $\mu $ cannot be obtained by the DA algorithm
with any single tiebreaking rule.\newline

Therefore, the negative transitivity of the priorities is important for
Remark 2 and Corollary 4.

\subsection{EADAM}

Since the mechanism in the class introduced above may not result in an SOSM
for $\succ $, several mechanisms that derive an SOSM for $\succ $ are
introduced by previous studies. Among them, Kesten (2010) and Tang and Yu
(2014) introduce a simple variant of EADAM to derive an SOSM for $\succ \in 
\mathcal{W}^{\left\vert S\right\vert }$.\footnote{%
Contrary to the DA mechanisms with tie-breaking, any SOSM mechanims are not
strategyproof (Abdulkadiro\u{g}lu et al. (2009)). However, several recent
studies show that the EADAM has several good incentive properties. See,
Cerrone et al. (2022), for the survey.}$^{,}$\footnote{%
In our personal communication, Professor Onur Kesten kindly told us that the
description of the EADAM variant introduced in Kesten (2010, Section V.D.)
was missing an additional requirement, corresponding to 2 for round $k\geq 1$
below, which was corrected in Tang and Yu (2014). Although we can use it
with the additional requirement, we have introduced the Tang and Yu (2014)
version of EADAM here. We appreciate the response from Professor Onur Kesten.%
}

\begin{description}
\item[Round $0$] Derive the SOSM for $\succ ^{\prime }\in \mathcal{E}\left(
\succ \right) $ e.g., via the DA algorithm.

\item[Round $k\geq 1$] 

\begin{enumerate}
\item Settle the matching at the underdemanded schools\footnote{%
School $s$ is said to be underdemanded at $\mu $ if there is no student $i$
such that $sP_{i}\mu \left( i\right) $.} at the resulting matching of Round $%
k-1$, and remove these schools and the students either matched with them or
not matched with any schools (matched with the \textquotedblleft null
school\textquotedblright\ in the words of Tang and Yu (2014)).

\item For each removed student $i$ and each remaining school $s$ that $i$
desires, if there is a remaining student $j$ such that $i\succ _{s}j$, then
remove $s$ from the preference of $j$.

\item Derive the SOSM of Round $k$ for the subproblem with only the
remaining schools and students whose preferences may be modified.
\end{enumerate}
\end{description}

The algorithm is terminated when all schools are removed.

Although Kesten (2010) and Tang and Yu (2014) only consider the case where $%
\succ \in \mathcal{W}^{\left\vert S\right\vert }$, Kitahara and Okumura
(2023) show that the result of the EADAM must be an SOSM for $\succ $ in the
case where $\succ \in \mathcal{P}^{\left\vert S\right\vert }$.

Since the result of the EADAM is dependent on the extension profile of $%
\succ $, let $EA\left( \succ ,\succ ^{\prime }\right) $ be the resulting
matching of the EADAM for $\succ $ and $\succ ^{\prime }\in \mathcal{E}%
\left( \succ \right) $, which is used in Round $0$.

Note that if the result of the EADAM at Round $0$, which is the SOSM for an
extension profile of $\succ ^{\prime }$, is also an SOSM for $\succ ,$ then
the result of the EADAM is equivalent to it. Otherwise, a Pareto improvement
is realized unless the matching is an SOSM for $\succ $ and hence $EA\left(
\succ ,\succ ^{\prime }\right) $ Pareto dominates the unique SOSM in $%
f^{\succ ^{\prime }}$. We have the following result.

\begin{corollary}
If $\succ \in \mathcal{P}^{\left\vert S\right\vert }$, then 
\begin{equation*}
f^{\succ }=\bigcup\limits_{\succ ^{\prime }\in \mathcal{E}\left( \succ
\right) }\left\{ EA\left( \succ ,\succ ^{\prime }\right) \right\} .
\end{equation*}
\end{corollary}

Dur et al. (2019, Proposition 4) show that if the priority profile domain is 
$\mathcal{W}^{\left\vert S\right\vert }$, then the EADAMs class (for weak
order priorities) is a member of the stable improvement cycle mechanisms
(SIC) class \`{a} la Erdil and Ergin (2008). Kitahara and Okumura (2021)
show that it is also true in a wider priority profile domain $\mathcal{P}%
^{\left\vert S\right\vert }$. Corollary 5 implies that if $\succ \in 
\mathcal{P}^{\left\vert S\right\vert }$, then the converse is also true;
that is, these classes are equivalent.

Next, let $d\left( \mu \right) $ be the set of matchings that Pareto
dominates $\mu $.

\begin{corollary}
Suppose $\succ \in \mathcal{P}^{\left\vert S\right\vert }$. For any $\mu \in 
\mathcal{S}^{\succ }$, 
\begin{equation*}
f^{\succ }\cap \left( d\left( \mu \right) \cup \left\{ \mu \right\} \right)
=\bigcup\limits_{\succ ^{\prime }\in \mathcal{E}\left( \succ \cup \left\{
\left. \left( i,j\right) \text{ }\right\vert \text{ }\mu \left( i\right)
=sP_{j}\mu \left( j\right) \right\} \right) }\left\{ EA\left( \succ ,\succ
^{\prime }\right) \right\} .
\end{equation*}
\end{corollary}

This implies that for any $\mu \in \mathcal{S}^{\succ }\setminus f^{\succ }$%
, we can have any SOSMs in $f^{\succ }$ that Pareto dominate $\mu $ by using
the SMO and the EADAM. Kitahara and Okumura (2021) show the same
characterization result of the SIC class.

Finally, we remark that if $\succ _{s}\in \mathcal{A\setminus P}$ for some $%
s\in S$, then any algorithm in the SIC and EADAM classes may not result in
an SOSM for $\succ $. See Kitahara and Okumura (2021, 2023) on the examples.

\subsection{Allowable Priority Violations}

We introduce the model of Dur et al. (2019), which is a generalization of
Kesten (2010). The model is defined as $G^{\prime }=\left( \succ ,C\right) $
such that $C:S\rightrightarrows I\times I$ be a correspondence, where $%
\left( i,j\right) \in C\left( s\right) $ means that the priority of $i$ for $%
s$ over $j$ is allowed to be violated. In the previous studies such as
Kesten (2010), Dur et al. (2019) and Kitahara and Okumura (2021), they
require some assumption on $C$. However, we do not require any assumption on 
$C$ here.

A matching is said to be \textbf{partially stable} for $\left( \succ
,C\right) $ if it is individually rational, non-wasteful, and for each $%
i,j\in I$ and $s\in S$, if $\mu \left( j\right) =s$, $sP_{i}\mu \left(
i\right) $ and $\left( i,j\right) \in \succ _{s}$, then $(i,j)\in C\left(
s\right) $. Let $\succ _{s}^{C}=\succ _{s}\setminus C\left( s\right) $ and $%
\succ ^{C}=\left( \succ _{s}^{C}\right) _{s\in S}$. The following result is
due to Kitahara and Okumura (2021, Remark 1).

\begin{remark}
A matching is partially stable for $\left( \succ ,C\right) $ if and only if
it is stable for $\succ ^{C}$.
\end{remark}

We have the following result.

\begin{proposition}
For any $\succ _{s}$ $\in \mathcal{A}$ and $C\left( s\right) $, $\succ
_{s}^{C}\in \mathcal{A}$. For any $\succ _{s}^{\prime }\in \mathcal{A},$
there is some $\succ _{s}$ $\in \mathcal{T}$ and some $C\left( s\right) $
such that $\succ _{s}^{C}=\succ _{s}^{\prime }$.
\end{proposition}

\textbf{Proof.} We show the first result. For any $C$, then $\succ
_{s}^{C}\subseteq \succ _{s}$. Thus, $\succ _{s}^{C}$ is also acyclic and
thus in $\mathcal{A}$.

We show the second result. By Proposition 1, for any $\succ _{s}^{\prime
}\in \mathcal{A}$, there is an extension of $\succ _{s}^{\prime }$ denoted
by $\succ _{s}^{\prime \prime }\in \mathcal{T}$. Let $\succ _{s}=\succ
_{s}^{\prime \prime }$ and $C\left( s\right) =\succ _{s}^{\prime \prime
}\setminus \succ _{s}^{\prime }$. Then, $\succ _{s}^{C}=\succ _{s}^{\prime }$%
. \textbf{Q.E.D.}\newline

Remark 3 and Proposition 2 imply that this model (even with $\succ _{s}$ $%
\in \mathcal{T}$) is essentially equivalent to the school choice model with $%
\succ \in \mathcal{A}^{\left\vert S\right\vert }$ introduced in this study.

If Assumption 2 of Kitahara and Okumura (2021), which is weaker than
Assumption 1 of Dur et al. (2019), is satisfied, then $\succ _{s}^{C}$ is
transitive. However, by Proposition 2, $\succ _{s}^{C}$ is acyclic but not
transitive in general even if $\succ _{s}\in \mathcal{T}$.

We consider an example where $\succ _{s}^{C}$ is acyclic but not transitive.
Suppose $\succ _{s}:$ $i_{1},i_{2},i_{3}$ and $C\left( s\right) =\left\{
\left( i_{1},i_{3}\right) \right\} $. Then, $\succ _{s}^{C}=$ $\left\{
\left( i_{1},i_{2}\right) ,\left( i_{2},i_{3}\right) \right\} $ is acyclic
but not transitive. For example, we consider the following scenario to
justify this situation. We assume that each student can decide whether to
waive a part of her/his priorities; that is, the \textit{all-or-nothing
property} considered by Dur et al. (2019) does not hold. Suppose that $i_{2}$
is a majority student and $i_{3}$ is a minority student. First, $i_{1}$
allows $\left( i_{1},i_{3}\right) $ to be violated, but does not allow $%
\left( i_{1},i_{2}\right) $ to be violated at $s$, because $i_{3}$ is a
minority student but $i_{2}$ is not. On the other hand, $i_{2}$ does not
allow any priority violation at $s$. Then, $C\left( s\right) =\left\{ \left(
i_{1},i_{3}\right) \right\} $ and thus $\succ _{s}^{C}$ is not transitive.

The following result is a generalization of Kitahara and Okumura (2021,
Proposition 2).

\begin{corollary}
If $\succ \in \mathcal{A}^{\left\vert S\right\vert }$, then there exists a
partially stable matching for $\left( \succ ,C\right) $.
\end{corollary}

This result is straightforward from Theorem 1 (Corollary 1), Remark 3 and
Proposition 2.

Finally, Dur et al. (2019) define that a matching is \textbf{constrained
efficient}\textit{\ }if it is partially stable and is not Pareto dominated
by any partially stable matching. If $\succ ^{C}\in \mathcal{P}^{\left\vert
S\right\vert }$, then we can computationally efficiently derive a
constrained efficient matching by using an algorithm in the SIC class or the
EADAMs class. However, as mentioned above, if $\succ _{s}^{C}\in \mathcal{%
A\setminus P}$ for some $s\in S$, then any algorithm in these classes may
not result in any constrained efficient matching. Therefore, by Proposition
2, in general, no computationally efficient mechanism to derive a
constrained efficient is known even if $\succ \in \mathcal{T}^{\left\vert
S\right\vert }$.

\section*{References}

\begin{description}
\item Abdulkadiro\u{g}lu, A., Pathak, P.A., Roth, A.E., 2009.
Strategy-proofness versus efficiency in matching with indifferences:
redesigning the NYC high school match.\ Am Econ Rev 99, 1954--1978.

\item Bossert, W., Suzumura, K. (2010) Consistency, Choice, and Rationality,
Harvard University Press.

\item Cerrone, C., Hermstr\"{u}wer, Y., Kesten, O. 2022. School Choice with
Consent: An Experiment, MPI Collective Goods Discussion Paper, No. 2022/2,
Available at SSRN: https://ssrn.com/abstract=4030661

\item Dur, U., Gitmez, A., Y\i lmaz, \"{O}. 2019. School choice under
partial fairness,\ Theor. Econ. 14(4), 1309-1346.

\item Erdil, A., Ergin, H. 2008. What's the matter with tie-breaking?
Improving efficiency in school choice,\ Am Econ Rev 98(3), 669--689.

\item Fishburn, P.C. 1970. Utility Theory for Decision Making, New York:
John Wiley and Sons.

\item Gale D., Shapley L. S. 1962. College admissions and the stability of
marriage. Am Math Mon 69(1):9--15

\item Kesten, O. 2010. School choice with consent. Quart J Econ
125(3):1297--1348

\item Kitahara, M., Okumura, Y. 2021. Improving efficiency in school choice
under partial priorities,\ Int J Game Theory 50, 971--987.

\item Kitahara, M., Okumura, Y. 2023. School Choice with Multiple
Priorities. Available at arXiv:2308.04780

\item Okumura, Y. 2016. A Stable and Pareto Efficient Update of Matching in
School Choice. Econ Lett 143, 111-113.

\item Okumura, Y. 2023. Consistent Linear Orders for Supermajority Rules.
Available at arXiv:2304.09419

\item Tang, Q., Yu, J., 2014. A new perspective on Kesten's school choice
with consent idea. J Econom Theory 154, 543--561.
\end{description}

\end{document}